\documentclass[10pt,letterpaper]{article}
\usepackage[top=0.85in,left=2.75in,footskip=0.75in]{geometry}

\usepackage{amsmath,amssymb}

\usepackage{changepage}

\usepackage[utf8x]{inputenc}

\usepackage{textcomp,marvosym}

\usepackage{cite}

\usepackage{nameref,hyperref}

\usepackage[right]{lineno}

\usepackage{microtype}
\DisableLigatures[f]{encoding = *, family = * }

\usepackage[table]{xcolor}

\usepackage{array}

\newcolumntype{+}{!{\vrule width 2pt}}

\newlength\savedwidth

\raggedright
\usepackage[aboveskip=1pt,labelfont=bf,labelsep=period,justification=raggedright,singlelinecheck=off]{caption}

\makeatletter
\renewcommand{\@biblabel}[1]{\quad#1.}
\makeatother

\usepackage{lastpage,fancyhdr,graphicx}
\usepackage{epstopdf}
\fancyheadoffset[L]{2.25in}
\fancyfootoffset[L]{2.25in}
\begin{document}
\vspace*{0.2in}

\begin{flushleft}
{\Large
\textbf\newline{Gravity model explained by the radiation model on a population landscape} 
}
\newline
\\
Inho Hong\textsuperscript{1},
Woo-Sung Jung\textsuperscript{1,2,3,4},
Hang-Hyun Jo\textsuperscript{3,1,5*}
\\
\bigskip
\textbf{1} Department of Physics, Pohang University of Science and Technology, Pohang 37673, Republic of Korea
\\
\textbf{2} Department of Industrial and Management Engineering, Pohang University of Science and Technology, Pohang 37673, Republic of Korea
\\
\textbf{3} Asia Pacific Center for Theoretical Physics, Pohang 37673, Republic of Korea
\\
\textbf{4} Department of Informatics, Indiana University Bloomington, Bloomington, IN 47408, USA
\\
\textbf{5} Department of Computer Science, Aalto University, Espoo FI-00076, Finland
\\
\bigskip

%
%





* hang-hyun.jo@apctp.org

\end{flushleft}
\section*{Abstract}
Understanding the mechanisms behind human mobility patterns is crucial to improve our ability to optimize and predict traffic flows. Two representative mobility models, i.e., radiation and gravity models, have been extensively compared to each other against various empirical data sets, while their fundamental relation is far from being fully understood. In order to study such a relation, we first model the heterogeneous population landscape by generating a fractal geometry of sites and then by assigning to each site a population independently drawn from a power-law distribution. Then the radiation model on this population landscape, which we call the radiation-on-landscape (RoL) model, is compared to the gravity model to derive the distance exponent in the gravity model in terms of the properties of the population landscape, which is confirmed by the numerical simulations. Consequently, we provide a possible explanation for the origin of the distance exponent in terms of the properties of the heterogeneous population landscape, enabling us to better understand mobility patterns constrained by the travel distance.


\section*{Introduction}\label{sec:intro}
For understanding the mechanisms of human mobility~\cite{Gonzalez2008Understanding, Song2010Modelling, Song2010Limits}, optimizing the mobility flows~\cite{Helbing2001Traffic}, and predicting the dynamics on mobility networks~\cite{Colizza2006Role, Balcan2009Multiscale, Brockmann2013Hidden}, a variety of mobility models have been extensively studied~\cite{Barbosa2018Human}, such as gravity model~\cite{Zipf1946P1}, intervening opportunities model~\cite{Stouffer1940Intervening}, and radiation model~\cite{Simini2012Universal}. Among these models, the gravity model has been widely used for predicting the traffic flows between populated areas. The gravity model predicts the traffic flow between an origin and a destination in terms of a simple formula, similar to Newton's gravity law, using populations of the origin and destination as well as the geographical distance between them~\cite{Zipf1946P1, Erlander1990Gravity, Barthelemy2011Spatial}. Precisely, the traffic from a site $i$ to another site $j$ is given by 
\begin{equation}
    \label{eq:gravity}
	T_{ij} \propto \frac{m_i m_j}{r_{ij}^{\gamma}},		
\end{equation}
where $m_i$ ($m_j$) denotes the population of site $i$ ($j$) and $r_{ij}$ is the distance between sites $i$ and $j$. The value of distance exponent $\gamma$ is found to range from $0.5$ to $3$ for several data sets~\cite{Barthelemy2011Spatial}.
This original gravity model and its variants have been applied to human mobility and transportation~\cite{Erlander1990Gravity, Jung2008Gravity, Barthelemy2011Spatial, Lenormand2012Universal, Simini2012Universal, Goh2012Modification, Masucci2013Gravity, Palchykov2014Inferring, Lee2014Matchmaker, Lee2015Relating, Park2018Generalized} ranging from the individual level~\cite{pappalardo2018data} to the international level~\cite{Balcan2009Multiscale}, and other datasets such as international trade~\cite{Bhattacharya2008International}, scientific collaboration~\cite{Pan2012World}, and mobile phone communication~\cite{Krings2009Urban, Palchykov2014Inferring} to name a few, mostly due to their simplicity. The gravity models nevertheless have limitations such as the absence of universality regarding the exponent estimation~\cite{Simini2012Universal}.

In order to overcome these limitations of the gravity models, Simini~\emph{et al.}~\cite{Simini2012Universal} recently suggested the radiation model, similar to the intervening opportunities model, that considers the opportunity for travelers rather than the distance traveled. By employing the radiation and absorption processes of particles, the radiation model describes the mobility patterns without any parameter estimation. Precisely, the traffic from a site $i$ to another site $j$ is given by 
\begin{equation}
    \label{eq:radiation}
	T_{ij} = T_{i} \frac{m_im_j}{(m_i+s_{ij})(m_i+s_{ij}+m_j)},
\end{equation}
where $T_i$ is the outgoing traffic from the site $i$ and $s_{ij}$ is the total population, except for the sites $i$ and $j$, within a circle centered at the site $i$ with radius $r_{ij}$~\cite{Simini2012Universal}. The radiation model has several advantages compared to the gravity model such as clear theoretical background, universality due to the absence of parameters to be estimated, and better prediction for long-distance travels, despite some unresolved issues like relatively poor predictability on short-distance travels~\cite{Masucci2013Gravity}. Moreover, the radiation model requires additional information on $T_i$, in contrast to the gravity model. The variants of the radiation and intervening opportunities models, e.g., a population-weighted opportunities model~\cite{Yan2014Universal} and a radiation model with an additional scaling exponent~\cite{Kang2015Generalized}, have also been studied. 

The radiation and gravity models have been compared with each other, often together with other mobility models, in terms of the predictability of mobility patterns observed in various empirical data sets~\cite{Masucci2013Gravity, Palchykov2014Inferring, Lenormand2016Systematic}. Here we raise a question: Beyond the comparison, can these radiation and gravity models be more fundamentally connected to each other? The possibility of such connection was briefly argued by Simini~\emph{et al.}~\cite{Simini2012Universal, Simini2013Human} such that the surrounding population $s_{ij}$ was assumed to be proportional to $r_{ij}^2$ in the case with the uniformly distributed population, and later to be proportional to $r_{ij}^{d_f}$ with the fractal dimension $d_f$ of the population. These assumptions lead to the asymptotic values of $\gamma=4$ and $2d_f$, respectively. However, the population landscape in reality can be characterized not only by a fractal geometry of populated areas or sites but also by a power-law distribution of the population at each site. In this paper, we first devise a population landscape model characterized both by a fractal dimension $d_f$ and by the power-law exponent $\beta$ of the population distribution, and then derive the distance exponent $\gamma$ as a function of $d_f$ and $\beta$ from the radiation model on our population landscapes, which we call the radiation-on-landscape (RoL) model. We also show that the distance exponent can vary according to the population sizes of origin and destination sites. These results shed light on the connection between gravity and radiation models. More importantly, we unveil the origin of the distance exponent in the gravity model in terms of the properties of the heterogeneous population landscape, provided that the radiation model is correct. Therefore we can better understand the mechanism behind the traffic flows constrained by the travel distance.

\section*{Results}

\subsection*{Modeling heterogeneous population landscapes}\label{sec:landscape}

As for the properties of heterogeneous population landscapes, we consider the fractal geometry of cities and the power-law distribution of their populations, both of which are well-known characteristics of human settlement. On the one hand, the fractal geometry suggested by Mandelbrot~\cite{Mandelbrot1967How} has been applied to the landscapes of human settlements in several states of the United States of America~\cite{sambrook2001fractal} and over the world~\cite{yook2002modeling}: The fractal dimension in those datasets is found to range from $1.4$ to $1.9$. The fractality has also been studied regarding the inner structures of cities~\cite{Batty1992Form, Batty1994Fractal, Shen2002Fractal} and their growth patterns~\cite{Makse1995Modelling, Benguigui2000When, Rybski2013Distanceweighted, Li2017Simple}. On the other hand, the power-law distribution of urban populations was presented in the classic paper by Zipf~\cite{Zipf1946P1} as well as in a number of recent studies~\cite{Rosen1980Size, Gabaix1999Zipfs, Soo2005Zipfs, Berry2012City, arshad2018zipf}. The power-law exponent of the population distribution of cities is found to have the value ranging from $1.7$ to $3$~\cite{Zipf1946P1, Rosen1980Size, Soo2005Zipfs, Clauset2009Powerlaw}. Despite the ongoing debate on whether populations are characterized by a power-law or a log-normal distribution~\cite{Eeckhout2004Gibrats, Berry2012City, Rozenfeld2011Area}, the power-law distribution would be still a reasonable assumption for model studies. 

For modeling the heterogeneous population landscape, we first generate a set of sites in a two-dimensional space with a fractal dimension $d_f$. Then we assign to each site $i$ the population $m_i$ independently drawn from $P(m)\sim m^{-\beta}$ with an exponent $\beta$, which will be called the population exponent. Note that the geometry of the sites can be implemented irrespective of the functional form of $P(m)$. In our work, we focus on the case in which the location and population of each site are fully uncorrelated with each other.

\begin{figure}[!t]
	\begin{adjustwidth}{-2.25in}{0in}
	\centering
	\includegraphics[width=19.5cm]{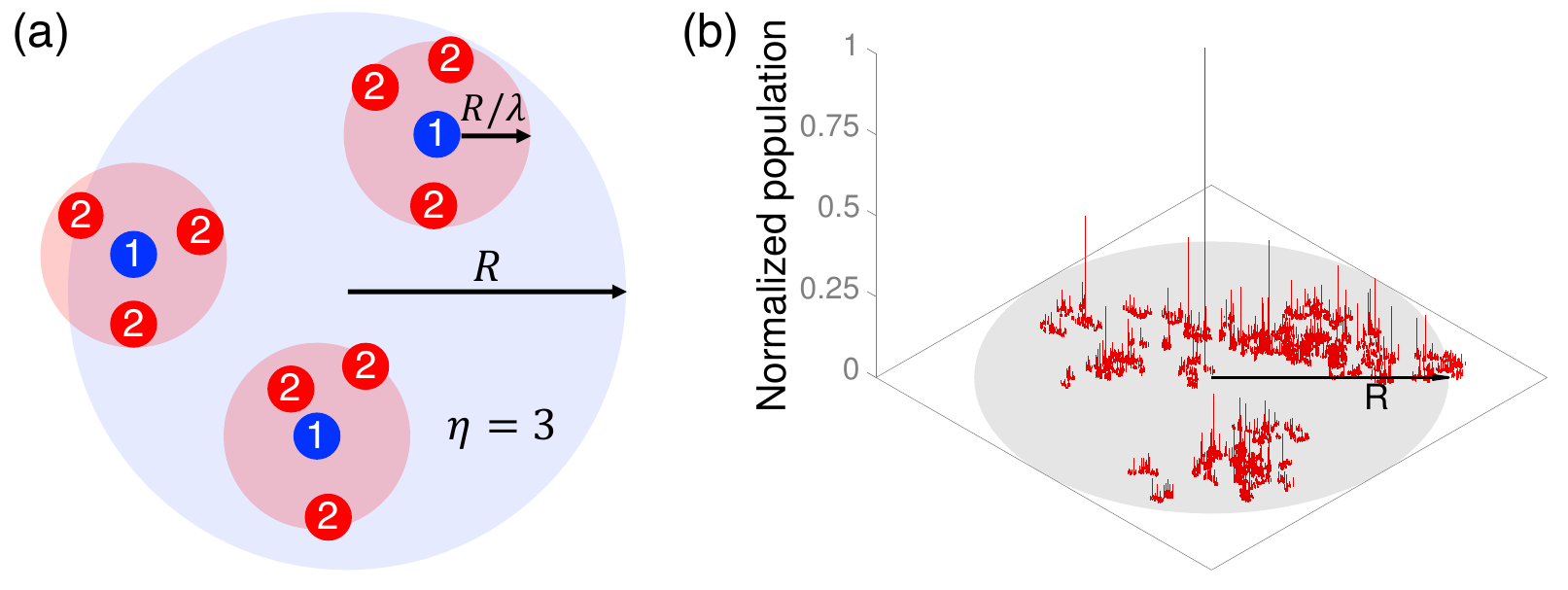}
    \caption{{\bf Modeling heterogeneous population landscapes.} (a) Schematic diagram of the Soneira-Peebles model in the two-dimensional space with $\eta=3$. The number in each circular symbol denotes the layer which it belongs to. The sites at the first layer (blue) are randomly placed within the circle with radius $R$. Similarly, the sites at the second layer (red) are randomly placed within the circles with radius $R/\lambda$. (b) An example of the generated population landscape using the Soneira-Peebles model with $\eta=2$, $\lambda=2^{1/1.5}$ (i.e., $d_f=1.5$), and $L=13$, and a population distribution $P(m)\sim m^{-\beta}$ with the population exponent $\beta=3$. The height in the vertical axis represents the normalized value of the population assigned to each site.}
	\label{fig:SP_model}
	\end{adjustwidth}
\end{figure}

In order to generate a fractal geometry of sites, we employ the Soneira-Peebles model~\cite{Soneira1978Computer}, originally developed for simulating the self-similar galaxy distribution. The model on the two-dimensional space iteratively locates sites within each circle centered at the site in the previous layer whose radius is decreasing as the layer deepens, see Fig~\ref{fig:SP_model}(a). Precisely, we consider a circle centered at the origin with radius $R$. Within this circle, $\eta>1$ sites are randomly placed in the first layer and each of these sites is assigned a circle with radius $R/\lambda$, with $\lambda>1$ denoting the contraction factor between layers. The same process is repeated until the depth of the layer reaches $L$, eventually leaving us with $\eta^{L}$ sites in the $L$th layer. Here $L$ denotes the number of layers. In our work, we consider the set of sites only in the last layer to find its fractal dimension as~\cite{Gospodinov2012Testing}
\begin{equation}
    \label{eq:fractal}
    d_{f}=\frac{\ln{\eta}}{\ln{\lambda}}.
\end{equation}
Once the set of $N=\eta^{L}$ sites with a fractal geometry is generated, we draw $N$ independent values from a population distribution $P(m)$ to randomly assign them to the sites. As for the population distribution we adopt the power-law distribution with the population exponent $\beta>1$:
\begin{equation}
    \label{eq:popul_distr}
    P(m)=(\beta-1)m_0^{\beta -1}m^{-\beta}\ \textrm{for}\ m\geq m_0.
\end{equation}
where $m_0$ is the lower bound of the population. We set $m_{0}=100$ to scale the population to a realistic size. Fig~\ref{fig:SP_model}(b) shows an example of the generated population landscape in the two-dimensional space using $\eta=2$, $\lambda=2^{1/1.5}$ (i.e., $d_f=1.5$), $L=13$, and $\beta=3$. The height in the vertical axis indicates the population assigned to each site. Although there exist many other modeling approaches for generating heterogeneous population landscapes~\cite{Witten1981DiffusionLimited, Makse1995Modelling, Schweitzer1998Estimation, Li2017Simple}, we have adopted the Soneira-Peebles model for the fractal geometry, mostly because the implementation of this model is efficient and scalable.

\subsection*{Connecting the radiation-on-landscape model to the gravity model}\label{sec:connect}

The connection between radiation and gravity models can be made by the observation that the surrounding population $s_{ij}$ of the radiation model in Eq~(\ref{eq:radiation}) might be correlated with the distance $r_{ij}$ of the gravity model in Eq~(\ref{eq:gravity}). The relation between $s_{ij}$ and $r_{ij}$ can be analytically derived in our population landscape model. Using this relation, the radiation model in our population landscape, i.e., the radiation-on-landscape (RoL) model, can be described by Eq~(\ref{eq:radiation}) but in terms of $r_{ij}$. By expanding the RoL model with respect to $r_{ij}$, one can derive the distance exponent $\gamma$ as a function of the fractal dimension $d_f$ and the population exponent $\beta$ of population landscapes.

\subsubsection*{Scaling behavior of surrounding population}\label{subsec:surround}

We first remind that the surrounding population $s_{ij}$ is defined as the total population, except for the sites $i$ and $j$, within a circle centered at the site $i$ with radius $r_{ij}$. Let us denote by $\Lambda_{ij}$ the set of sites, except for $i$ and $j$, within a circle centered at the site $i$ with radius $r_{ij}$, and the number of sites in $\Lambda_{ij}$ is denoted by $n_{ij}$. In a $d_f$-dimensional space, one can write as
\begin{equation}
	\label{eq:number_radius}
    n_{ij}=cr_{ij}^{d_f},
\end{equation}
with a coefficient $c$. The surrounding population is written as
\begin{equation}
    \label{eq:surround}
    s_{ij}= \sum_{l\in \Lambda_{ij}} m_l = \sum_{k=1}^{n_{ij}} m_k.
\end{equation}
where $m_k$ denotes the population of the $k$th populated site in $\Lambda_{ij}$, such that $m_1\geq m_2\geq\cdots\geq m_{n_{ij}}$. As all $m_l$s are statistically independent of each other, one can relate $m_k$ with its rank $k$ using $P(m)$ as~\cite{Sornette2006Critical}
\begin{equation}
    \frac{k}{n_{ij}}=\int_{m_k}^\infty P(m)dm.
\end{equation}
From Eq~(\ref{eq:popul_distr}) we have
\begin{equation}
    m_k=m_0 \left(\frac{k}{n_{ij}}\right)^{-1/(\beta-1)},
\end{equation}
where we note that $\beta>1$, leading to
\begin{equation}
	\label{eq:s_approx}
    s_{ij} \approx \int_{1}^{n_{ij}} m_k dk =\left\{
        \begin{tabular}{ll} 
            $\frac{\beta-1}{\beta-2}m_0 \left(n_{ij} - n_{ij}^{1/(\beta-1)}\right)$ & for $\beta\neq 2$,\\
            $m_0 n_{ij}\ln n_{ij}$ & for $\beta=2$.\\
        \end{tabular}\right.
\end{equation}
Therefore one gets
\begin{equation}
	\label{eq:s_approx_r}
    s_{ij} \approx \left\{
        \begin{tabular}{ll} 
            $\frac{\beta-1}{\beta-2}m_0 \left(cr_{ij}^{d_f} - c^{1/(\beta-1)}r_{ij}^{d_f/(\beta-1)}\right)$ & for $\beta\neq 2$,\\
            $m_0 cr_{ij}^{d_f}\ln (cr_{ij}^{d_f})$ & for $\beta=2$,\\
        \end{tabular}\right.
\end{equation}
where we have used Eq.~(\ref{eq:number_radius}). When $\beta>2$, the term of $r_{ij}^{d_f}$ dominates $s_{ij}$ for large $r_{ij}$, while the term of $r_{ij}^{d_f/(\beta-1)}$ does for $\beta<2$. Therefore, we obtain the scaling relation between $s_{ij}$ and $r_{ij}$ for large $r_{ij}$:
\begin{equation}
    \label{eq:alpha}
    s_{ij}\sim r_{ij}^\alpha,
\end{equation}
with
\begin{equation}
    \label{eq:sij}
    \alpha = \left\{\begin{tabular}{ll}
        $d_{f}/(\beta-1)$ & for $\beta<2$,\\
        $d_{f}$ & for $\beta > 2$.
    \end{tabular}\right.
\end{equation}

\subsubsection*{Expansion of the RoL model}\label{subsec:distance}

The relation between $s_{ij}$ and $n_{ij}$ in Eq~(\ref{eq:s_approx}), together with the relation between $n_{ij}$ and $r_{ij}$ in Eq~(\ref{eq:number_radius}), allows us to rewrite the radiation model in terms of $r_{ij}$, i.e., the RoL model. From Eq~(\ref{eq:radiation}) we define the travel probability as
\begin{equation}
    \label{eq:pij}
    p_{ij} \equiv \frac{T_{ij}}{T_i} = \frac{m_im_j}{(m_i+s_{ij})(m_i+s_{ij}+m_j)},
\end{equation}
and the rescaled travel probability as
\begin{equation}
    \label{eq:rescaled}
    \frac{p_{ij}}{m_im_j} = \frac{1}{(m_i+s_{ij})(m_i+s_{ij}+m_j)}.
\end{equation}
For the expansion, we consider three cases: (i) $m_i, m_j\ll s_{ij}$, (ii) $m_i\ll s_{ij}\ll m_j$, and (iii) $m_i\gg s_{ij}$. 

(i) If $m_i, m_j \ll s_{ij}$, the rescaled travel probability is expanded as
\begin{equation}
    \label{eq:rescaled_case1}
    \frac{p_{ij}}{m_im_j} \approx s_{ij}^{-2}\left[1-(2m_i+m_j)s_{ij}^{-1}+\mathcal{O}\left( \frac{m_i^2}{s_{ij}^2}\right)\right].
\end{equation}
Here we find the leading term of $s_{ij}^{-2}\sim r_{ij}^{-2\alpha}$ from Eq.~(\ref{eq:alpha}), leading to
\begin{equation}
	\label{eq:nd_gravity}
    \frac{p_{ij}}{m_im_j}\sim \left\{\begin{tabular}{ll}
            $r_{ij}^{-2d_{f}/(\beta-1)}$ & for $\beta<2$,\\
            $r_{ij}^{-2d_{f}}$ & for $\beta > 2$.
    \end{tabular}\right.
\end{equation}
This scaling form of the distance dependence enables us to compare our RoL model to the gravity model in Eq~(\ref{eq:gravity}):
\begin{equation}
    \frac{T_{ij}}{m_{i}m_{j}}\sim r_{ij}^{-\gamma}.
\end{equation}
By comparing the distance dependence of the RoL and gravity models, we obtain the distance exponent $\gamma$ as a function of the fractal dimension $d_f$ and the population exponent $\beta$:
\begin{equation}
    \gamma=2\alpha= \left\{\begin{tabular}{ll}
        $2d_{f}/(\beta-1)$ & for $\beta<2$,\\
        $2d_{f}$ & for $\beta > 2$.
    \end{tabular}\right.
    \label{eq:distance_exponent_case1}
\end{equation}
Note that the result of $\gamma=2d_f$ has been suggested in a previous work~\cite{Simini2013Human}. 

(ii) If $m_i\ll s_{ij}\ll m_j$, one gets
\begin{equation}
    \label{eq:rescaled_case2}
    \frac{p_{ij}}{m_im_j} \approx \frac{s_{ij}^{-1}}{m_j} \left[1+ \left(\frac{m_i^2}{m_j}-m_i\right)s_{ij}^{-1} -\frac{s_{ij}}{m_j} + \mathcal{O}\left( \frac{m_i^2}{s_{ij}^2} + \frac{s_{ij}^2}{m_j^2} \right)\right].
\end{equation}
From the leading term of $s_{ij}^{-1}\sim r_{ij}^{-\alpha}$, we obtain
\begin{equation} 
    \gamma=\alpha= \left\{\begin{tabular}{ll}
        $d_{f}/(\beta-1)$ & for $\beta<2$,\\
        $d_{f}$ & for $\beta > 2$.
    \end{tabular}\right.
    \label{eq:distance_exponent_case2}
\end{equation}

(iii) Finally, if $m_i\gg s_{ij}$, one has
\begin{equation}
    \label{eq:rescaled_case3}
    \frac{p_{ij}}{m_im_j} \approx \frac{1}{m_i(m_i+m_j)} \left[1- \frac{2m_i+m_j}{m_i(m_i+m_j)} s_{ij} + \mathcal{O}\left(
            \frac{s_{ij}^2}{m_i^2}
        \right) \right],
\end{equation}
irrespective of $m_j$. Since the leading term $\frac{1}{m_i(m_i+m_j)}$ is independent of $r_{ij}$, we have 
\begin{equation} 
    \gamma=0.
    \label{eq:distance_exponent_case3}
\end{equation}
However, the subleading terms are still functions of $r_{ij}$, leading to a weak distant-dependent behavior of the rescaled travel probability.

From the above analysis, it is remarkable to find how the distance exponent $\gamma$ can vary according to the population sizes of origin and destination sites, i.e., $m_i$ and $m_j$, respectively. This strongly implies that a given data set does not necessarily have to be characterized by the single value of the distance exponent. In reality, travelers from small towns may have different reasons for selecting their destinations, hence different travel distances, than those from big cities; the population size of the destination can also affect the traveling behaviors.

We provide an intuitive explanation for our results in Eqs~(\ref{eq:distance_exponent_case1}) and~(\ref{eq:distance_exponent_case2}). We remind that in the gravity model, the distance exponent $\gamma$ plays a role of spatial cost in determining the traffic flows because the larger $\gamma$ leads to the stronger dependence of the traffic flows on the distance. Let us consider a job-seeking situation as in the original radiation model~\cite{Simini2012Universal}. Since the number of cities is proportional to $r^{d_{f}}$, a higher-dimensional geometry with a larger $d_f$ would provide more opportunities in the same range of $r$ from the origin. It implies that a job-seeker can find a job at a closer city and does not need to travel farther in a higher-$d_{f}$ space, leading to a larger $\gamma$. Dependency of $\gamma$ on the heterogeneity of the population distribution can also be understood with the job-seeking example. In the original radiation model, a place with the larger population provides more opportunities, and a job-seeker finds a job at the closest city providing the better opportunity than the origin. For example, let us consider a homogeneous case with $10$ medium-sized cities with two workplaces per city, which can be contrasted to a heterogeneous case with one extremely large city with $11$ workplaces and nine small cities with one workplace per each. Then the job seekers in the homogeneous case tend to travel to any other cities providing a little better opportunities, implying a smaller $\gamma$. In contrast, the job seekers in the heterogeneous case tend to travel only to the extremely large city and do not have to travel farther than that city, implying a larger $\gamma$. Since the smaller $\beta$ implies a more heterogeneous population distribution, one can relate the smaller $\beta$ to the larger $\gamma$, closing our arguments for Eqs~(\ref{eq:distance_exponent_case1}) and~(\ref{eq:distance_exponent_case2}).

\subsection*{Numerical validation}\label{sec:validate}

We numerically test the analytic results using the heterogeneous population landscapes described in Fig~\ref{fig:SP_model}. We generate $100$ different population landscapes with the same parameter set of $\eta=2$, $\lambda=2^{1/1.5}$ (i.e., $d_f=1.5$), $R=1$, and $L=13$, then assign to the sites the populations drawn from $P(m)\propto m^{-\beta}$ in Eq~(\ref{eq:popul_distr}). We also set the upper bound of $m_i$ as $10^7$. Once the population landscapes are generated, one can calculate for every pair of sites $i$ and $j$ the distance $r_{ij}$, the number of sites for the surrounding population $n_{ij}$, the surrounding population $s_{ij}$, and the travel probability $p_{ij}$ using the following Eq~(\ref{eq:radiation_finite}) for the finite system. The travel probability for the finite system~\cite{Masucci2013Gravity} is given by
\begin{equation}
    \label{eq:radiation_finite}
    p_{ij}\equiv \frac{T_{ij}}{T_i} = \frac{1}{1-\frac{m_i}{M}}\frac{m_im_j}{(m_i+s_{ij})(m_i+s_{ij}+m_j)},
\end{equation}
where $M\equiv \sum_i m_i$ denotes the total population in the system. As almost all $m_i$s are much smaller than $M$, our analytic results remain valid.

\begin{figure}[!t]
	\begin{adjustwidth}{-2.25in}{0in}
	\centering
	\includegraphics[width=19cm]{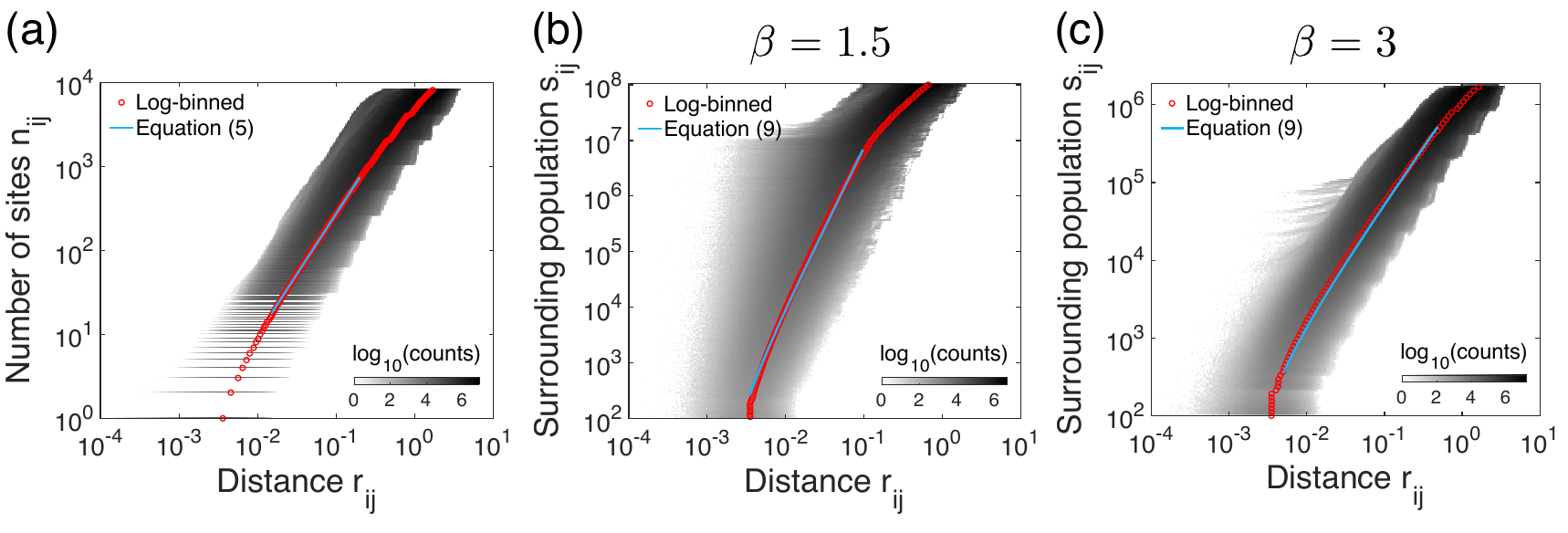}
    \caption{{\bf Properties of the heterogeneous population landscapes.} (a) Numerical validation of the scaling relation between $n_{ij}$ and $r_{ij}$ in Eq~(\ref{eq:number_radius}) for the fractal geometry of sites generated using Soneira-Peebles model with $\eta=2$, $\lambda=2^{1/1.5}$ (i.e., $d_f=1.5$), $R=1$, and $L=13$, averaged over $100$ different landscapes. (b,c) Numerical validation of the analytic relation between $s_{ij}$ and $r_{ij}$ in Eq~(\ref{eq:s_approx}), together with Eq~(\ref{eq:number_radius}), on the same fractal geometry of sites as in (a), but also with $P(m)\sim m^{-\beta}$ for the values of $\beta=1.5$ (b) and of $\beta=3$ (c), respectively. For each gray-colored heat map, the darker color implies more pairs of sites around the point $(r_{ij}, n_{ij})$ or $(r_{ij}, s_{ij})$. The log-binned curve (red circles) of the heat map is compared to the corresponding equation (light blue curve).
    }
	\label{fig:r_vs_s}
	\end{adjustwidth}
\end{figure}

\begin{figure}[!t]
	\begin{adjustwidth}{-2.25in}{0in}
	\centering
	\includegraphics[width=19.5cm]{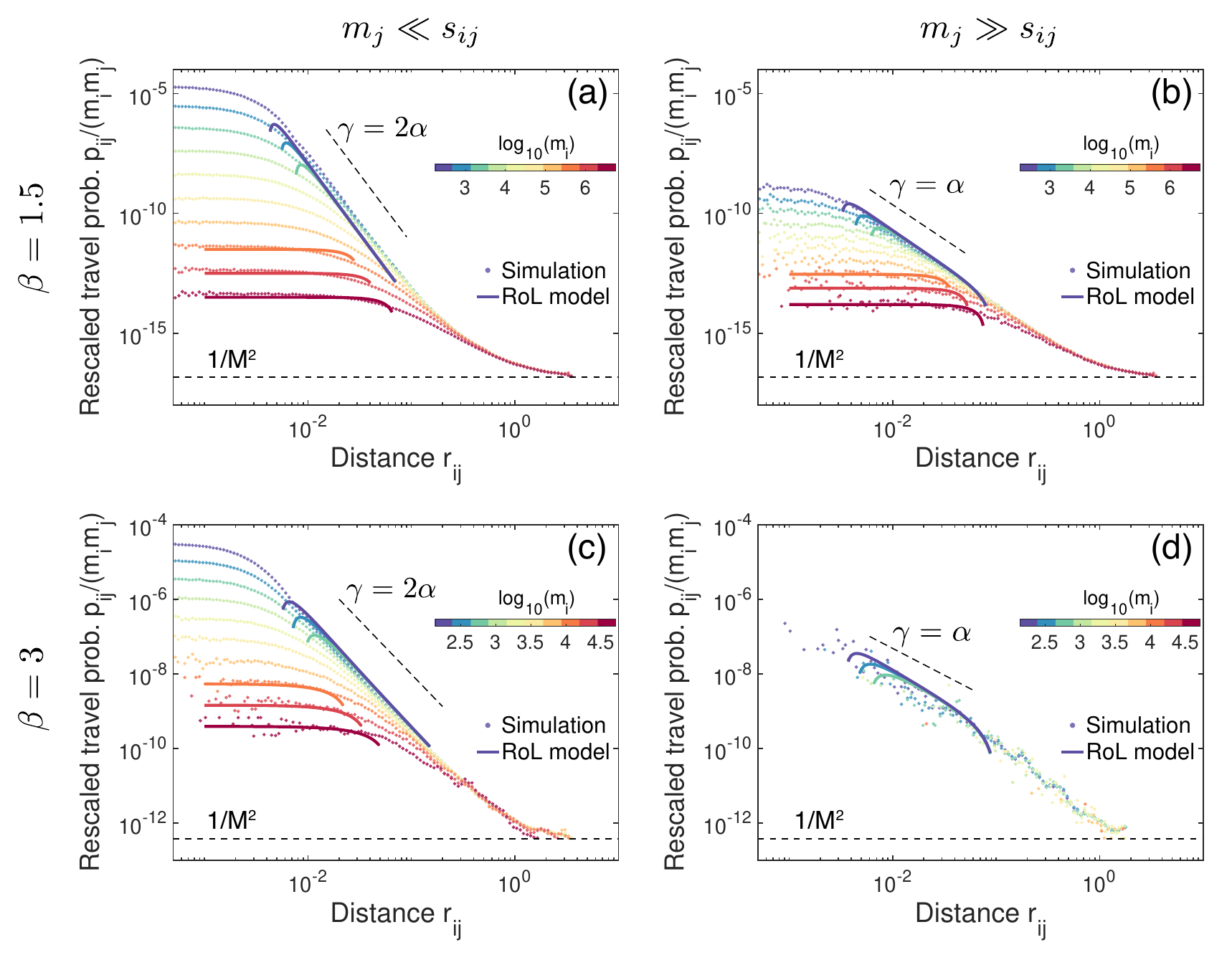}
    \caption{{\bf Numerical validation of the expanded forms of the radiation-on-landscape (RoL) model.} The expanded forms of the rescaled travel probabilities in Eqs~(\ref{eq:rescaled_case1}),~(\ref{eq:rescaled_case2}), and~(\ref{eq:rescaled_case3}) (solid curves) are tested against the numerical results using $p_{ij}$ in Eq~(\ref{eq:radiation_finite}) calculated on the same population landscapes used in Fig~\ref{fig:r_vs_s} (symbols), for the values of $\beta=1.5$ (top) and $3$ (bottom), respectively. For clearer visualization, we show only the curves of the rescaled travel probability for the group of the smallest $m_j$s (left) and those for the group of the largest $m_j$s (right), but for all groups of $m_i$ in each panel. The analytic results of the distance exponent $\gamma$ in Eqs~(\ref{eq:distance_exponent_case1}),~(\ref{eq:distance_exponent_case2}), and~(\ref{eq:distance_exponent_case3}) are also plotted by black dashed lines for comparison.}
	\label{fig:p_vs_r}
	\end{adjustwidth}
\end{figure}

\subsubsection*{Surrounding population}

The results of $(r_{ij}, n_{ij})$ for all possible pairs of sites $i$ and $j$ are depicted as a heat map in Fig~\ref{fig:r_vs_s}(a), from which we estimate the fractal dimension $\hat{d_f}\approx 1.44\pm0.07$ and the coefficient $\hat c\approx 7.55\times10^{3}$ in Eq~(\ref{eq:number_radius}). Here the scaling behavior is observed in the intermediate range of $r_{ij}$. The lower bound of this range is related to the smallest length scale, i.e., $R/\lambda^L\approx 10^{-2}$ for the parameter values used, while the upper bound is related to the largest length scale, which is trivially $R=1$. 

Similarly, from the results of $(r_{ij}, s_{ij})$ for all possible pairs of sites $i$ and $j$, we get the heat map for a few values of $\beta$, as shown in Fig~\ref{fig:r_vs_s}(b,c). When log-binned, it gives the curve of $s_{ij}$ as a function of $r_{ij}$, which turns out to be comparable to the analytic result in Eq~(\ref{eq:s_approx}) when using estimated values of $\hat{d_f}$ and $\hat{c}$ for both cases with $\beta<2$ and $\beta>2$. Accordingly, the scaling relation between $\alpha$, $d_f$, and $\beta$ in Eq~(\ref{eq:sij}) is also validated.

\subsubsection*{Rescaled travel probability}

Next, we test the validity of the expanded forms of rescaled travel probabilities in Eqs~(\ref{eq:rescaled_case1}),~(\ref{eq:rescaled_case2}), and~(\ref{eq:rescaled_case3}), by comparing them to the numerical results on the generated population landscapes using Eq~(\ref{eq:radiation_finite}). In particular, for studying the effects of origin and destination populations on the scaling behavior of the rescaled travel probability, the sites are decomposed into $10$ groups according to their population sizes, denoted by $G_v$ for $v=1,\cdots,10$. Then all pairs of origin and destination sites can be decomposed into $100$ groups of pairs, such that $G_{vw}=\{(i,j)|i\in G_v\ \textrm{and}\ j\in G_w\}$ for $v,w=1,\cdots,10$. For each group of pairs, say $G_{vw}$, we calculate the rescaled travel probabilities for all pairs in $G_{vw}$ using $p_{ij}$ in Eq~(\ref{eq:radiation_finite}) to obtain a heat map for $(r_{ij}, \frac{p_{ij}}{m_im_j})$ (not shown). By log-binning the heat map, one gets the curve of the rescaled travel probability as a function of $r_{ij}$ for each $G_{vw}$, as shown in Fig~\ref{fig:p_vs_r}. We find that these numerical results are in good agreement with the expanded forms of rescaled travel probabilities for $m_i,m_j\ll s_{ij}$ in Eq~(\ref{eq:rescaled_case1}), for $m_i\ll s_{ij}\ll m_j$ in Eq~(\ref{eq:rescaled_case2}), and for $m_i\gg s_{ij}$ in Eq~(\ref{eq:rescaled_case3}), respectively. Accordingly, the scaling relations between $\gamma$ and $\alpha$, i.e., the scaling relations between $\gamma$, $d_f$, and $\beta$ in Eqs~(\ref{eq:distance_exponent_case1}),~(\ref{eq:distance_exponent_case2}), and~(\ref{eq:distance_exponent_case3}) are also validated. This implies that the distance exponent $\gamma$ can vary according to the population sizes of origin and destination sites, even in the same population landscape. Here we remark that a recent empirical study showed that the origin and destination populations affect the travel patterns, whereas the distance exponent has been assumed to be the same irrespective of the populations~\cite{curiel2018gravity}.

\begin{figure}[!t]
	\begin{adjustwidth}{-2.25in}{0in}
	\centering
	\includegraphics[width=19.5cm]{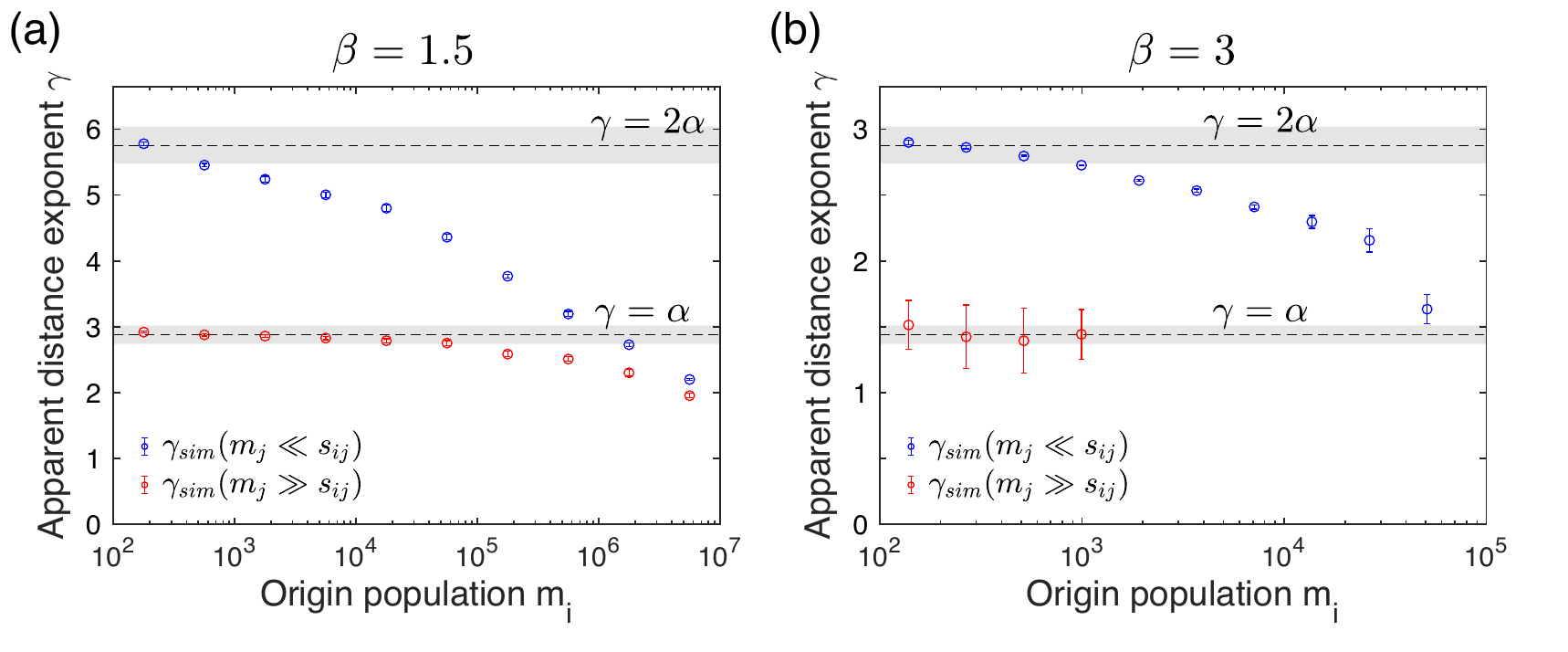}
    \caption{{\bf Behaviors of the apparent distance exponent $\gamma_{vw}$ according to the origin and destination populations.} We estimate the values of the apparent distance exponent $\gamma_{vw}$, defined by Eq~(\ref{eq:gamma_sim}), from the numerical curves of the rescaled travel probability shown in Fig~\ref{fig:p_vs_r} for the values of $\beta = 1.5$ (a) and $3$ (b), respectively. These values (symbols) are compared to analytic values for the limiting cases, i.e., $\gamma=2\alpha$ for $m_i,m_j\ll s_{ij}$ and $\gamma=\alpha$ for $m_i\ll s_{ij}\ll m_j$, which are plotted by black dashed lines with gray shadows denoting $\gamma\pm\sigma_{\gamma}$. Here $\sigma_\gamma$ is determined using the standard deviation of the estimated $\hat d_f$.}
	\label{fig:gamma_sim}
	\end{adjustwidth}
\end{figure}

We remark that the number of pairs of highly populated sites is in general much lower than those of other cases, so that the corresponding curves of the rescaled travel probability tend to be more fluctuating or even apparently missing, e.g., in the case with the groups of large $m_i$ and $m_j$ for $\beta=3$ in Fig~\ref{fig:p_vs_r}(d). Except for this case, we generically observe clear scaling behaviors of the rescaled travel probability in the intermediate range of $r_{ij}$. In addition, the curves are found to saturate to a constant for sufficiently small $r_{ij}$, whereas for sufficiently large $r_{ij}$, these curves converge to eventually approach the lower bound of the rescaled travel probability, $\frac{1}{M^2}$. These findings can be explained by Eq~(\ref{eq:radiation_finite}): On the one hand, for sufficiently small $r_{ij}$, $s_{ij}$ becomes negligible as there would be only few or even no sites in the surrounding area between $i$ and $j$. Thus, the rescaled travel probability becomes independent of $r_{ij}$ as $\frac{p_{ij}}{m_i m_j} \approx \frac{1}{m_i(m_i + m_j)}$. On the other hand, if $r_{ij}$ becomes sufficiently large, $s_{ij}$ approaches the total population $M$, irrespective of $m_i$ and $m_j$. This is why all curves converge and eventually approach the lower bound of the rescaled travel probability as $\frac{p_{ij}}{m_i m_j} \approx \frac{1}{M^2}$.

Finally, we discuss the generic behavior of the distance exponent according to the population sizes of origin and destination sites. We first point out that the scaling relations in Eqs~(\ref{eq:distance_exponent_case1}),~(\ref{eq:distance_exponent_case2}), and~(\ref{eq:distance_exponent_case3}) have been derived in the limiting cases of $m_i$ and $m_j$. Therefore, these results cannot be simply applied to the scaling behaviors observed for the cases with intermediate ranges of $m_i$ and $m_j$. For these cases, one can estimate the apparent distance exponent $\gamma_{vw}$ based on the assumption of the simple scaling form as
\begin{equation}
    \frac{p_{ij}}{m_im_j}\sim r_{ij}^{-\gamma_{vw}}
    \label{eq:gamma_sim}
\end{equation}
for each group of pairs $G_{vw}$. It is found that the value of $\gamma_{vw}$ appears to be continuously varying according to the origin population $m_i$  for the smallest and the largest groups of $m_j$, as depicted in Fig~\ref{fig:gamma_sim}. For example, when $m_j\ll s_{ij}$, the value of $\gamma_{vw}$ is $\approx 2\alpha$ for $m_i\ll s_{ij}$, and then it continuously decreases as $m_i$ increases. We also find the clear dependency of $\gamma_{vw}$ on the destination population $m_j$ for a given $m_i$.

\section*{Conclusion}\label{sec:conclusion}

Although two representative mobility models, i.e., gravity and radiation models, have been compared to each other against the empirical traffic data sets~\cite{Masucci2013Gravity, Palchykov2014Inferring, Lenormand2016Systematic}, the more fundamental connection between these models has been far from being fully understood. In order to study such a connection in a realistic population landscape, we first model the heterogeneous population landscape by assuming a fractal geometry of sites and the population at each site following a power-law distribution. Then the radiation model on such population landscapes, namely, the radiation-on-landscape (RoL) model, can be written in terms of the distance between two sites. By expanding the rescaled travel probability in the RoL model and comparing it to the gravity model, we derive the distance exponent in the gravity model as a function of the fractal dimension and the population exponent of the population distribution. We also find that this distance exponent can vary according to the population sizes of origin and destination sites. These analytic expectations are confirmed by numerical simulations on our population landscapes. Consequently, we could connect two representative mobility models, and more importantly, the origin of the distance exponent could be related to the properties of the heterogeneous population landscape as well as the population sizes of origin and destination sites. Therefore we can better understand the mechanism behind the traffic flows constrained by the travel distance. In particular, the effects of the populations of origins and destinations on the distance exponent can be empirically studied as a future work.

In our work we have assumed that the location and population of each site are fully uncorrelated with each other, while there might be some correlations between them in reality. One can study the effects of spatial correlations, e.g., by the positively correlated populations at close sites, on the traffic flows and their characteristic distance exponent. In addition, as for the functional form of the population distribution, one can adopt other functional forms than the power law, such as the log-normal distribution given by Gibrat's law~\cite{Eeckhout2004Gibrats}.

Finally, we remark that the mass term $m_i$ in many mobility models has been used to denote the population at the site, while it can be interpreted as other sources of attraction of sites, e.g., each site's traffic volume~\cite{Masucci2013Gravity}, economic indicator~\cite{Bhattacharya2008International}, communication volume~\cite{Palchykov2014Inferring}, and citations~\cite{Pan2012World}. Indeed, the diverse values of distance exponent have been observed according to the modes of transportation, geographical regions, and granularities~\cite{Barthelemy2011Spatial}. Considering our above findings on the mass dependency of the traffic flows, it is of crucial importance to empirically and theoretically relate various observables attributed to the site for better understanding of the human mobility.

\section*{Acknowledgments}
W.-S.J. was supported by Basic Science Research Program through the National Research Foundation of Korea (NRF) funded by the Ministry of Education (2016R1D1A1B03932590). H.-H.J. was supported by Basic Science Research Program through the National Research Foundation of Korea (NRF) funded by the Ministry of Education (NRF-2018R1D1A1A09081919).

\nolinenumbers

\bibliographystyle{plos2015}

\end{document}